\def\papertitle{Differentiable All-pass Filters for Phase
Response Estimation and Automatic Signal
Alignment}
\def\paperauthorA{Anders R. Bargum}
\def\paperauthorB{Stefania Serafin}
\def\paperauthorC{Cumhur Erkut}
\def\paperauthorD{Julian D. Parker}

\documentclass[twoside,a4paper]{article}
\usepackage{arxiv}
\usepackage{etoolbox}
\usepackage{amsmath,amssymb,amsfonts,amsthm}
\usepackage{euscript}
\usepackage[T1]{fontenc}
\usepackage[utf8]{inputenc}
\usepackage{ifpdf}
\usepackage[english]{babel}
\usepackage{caption}
\usepackage{subfig} 
\usepackage{color}
\usepackage{multirow}
\usepackage{caption} 
\input glyphtounicode
\newcounter{numauth}
\newcounter{listcnt}
\newcommand\authcnt[1]{\ifdefined#1 \stepcounter{numauth} \fi}

\newcommand\addauth[1]{
\ifdefined#1 
\stepcounter{listcnt}
\ifnum \value{listcnt}<\value{numauth}
\appto\authorslist{, #1}
\else
\appto\authorslist{~and~#1}
\fi
\fi}
\authcnt{\paperauthorB}
\authcnt{\paperauthorC}
\authcnt{\paperauthorD}
\authcnt{\paperauthorE}
\authcnt{\paperauthorF}
\authcnt{\paperauthorG}
\authcnt{\paperauthorH}
\authcnt{\paperauthorI}
\authcnt{\paperauthorJ}
\def\authorslist{\paperauthorA}
\addauth{\paperauthorB}
\addauth{\paperauthorC}
\addauth{\paperauthorD}
\addauth{\paperauthorE}
\addauth{\paperauthorF}
\addauth{\paperauthorG}
\addauth{\paperauthorH}
\addauth{\paperauthorI}
\addauth{\paperauthorJ}

\usepackage{times}

\newif\ifpdf
\ifx\pdfoutput\relax
\else
   \ifcase\pdfoutput
      \pdffalse
   \else
      \pdftrue
\fi

\ifpdf 
  \usepackage[pdftex,
    pdftitle={\papertitle},
    pdfauthor={\authorslist},
    pdfsubject={Proceedings of the 26th International Conference on Digital Audio Effects (DAFx23)},
    colorlinks=false, 
    bookmarksnumbered, 
    pdfstartview=XYZ 
  ]{hyperref}
  \usepackage[pdftex]{graphicx}
\else 
  \usepackage[dvips]{epsfig,graphicx}
  \usepackage[dvips,
    pdftitle={\papertitle},
    pdfauthor={\authorslist},
    pdfsubject={Proceedings of the 26th International Conference on Digital Audio Effects (DAFx23)},
    colorlinks=false, 
    bookmarksnumbered, 
    pdfstartview=XYZ 
  ]{hyperref}
\fi
\usepackage[hypcap=true]{caption}
\usepackage{comment}
\usepackage{cite}
\title{\papertitle}

\author{\paperauthorA*, \paperauthorB, \paperauthorC \\
{\href{https://melcph.create.aau.dk/}{Multi Sensory Experience Lab}, Aalborg University, Copenhagen, Denmark} \\
{\tt \href{mailto:arba@create.aau.dk}{arba@create.aau.dk}, \href{mailto:sts@create.aau.dk}{sts@create.aau.dk},\href{mailto:cer@create.aau.dk}{cer@create.aau.dk}} \\
{\bf \paperauthorD} \thanks{\textit{Collaboration done while interning/employed at Native Instruments. Accepted for publication in Proc. DAFX'23, Copenhagen, Denmark, September 2023}} \\
{{\tt \href{mailto:firstname.lastname@cantab.net}{firstname.lastname@cantab.net}}}}

\begin{document}
\ifpdf 
  \DeclareGraphicsExtensions{.png,.jpg,.pdf}
\else  
  \DeclareGraphicsExtensions{.eps}
\fi

\maketitle

\begin{abstract}
Virtual analog (VA) audio effects are increasingly based on neural networks and deep learning frameworks. Due to the underlying black-box methodology, a successful model will learn to approximate the data it is presented, including potential errors such as latency and audio dropouts as well as non-linear characteristics and frequency-dependent phase shifts produced by the hardware. The latter is of particular interest as the learned phase-response might cause unwanted audible artifacts when the effect is used for creative processing techniques such as dry-wet mixing or parallel compression. To overcome these artifacts we propose differentiable signal processing tools and deep optimization structures for automatically tuning all-pass filters to predict the phase response of different VA simulations, and align processed signals that are out of phase. The approaches are assessed using objective metrics while listening tests evaluate their ability to enhance the quality of parallel path processing techniques. Ultimately, an over-parameterized, BiasNet-based, all-pass model is proposed for the optimization problem under consideration, resulting in models that can estimate all-pass filter coefficients to align a dry signal with its affected, wet, equivalent.
\end{abstract}

\section{Introduction}
\label{sec:intro}
Digital simulations of analog audio equipment like tape machines, pre-amplifiers and distortion pedals remain in demand due to the hardware's rich history and unique sonic characteristics. With the increase in
computational power, the deep learning approach to machine learning has proven useful for simulating virtual analog (VA) black-box models and has in several publications been applied as the main technique for approximating the output response of analog audio systems \cite{chowdhury, damskagg2, Damskgg2019RealTimeMO, rnn_vela, Bergner}. In \cite{damskagg2} and \cite{Damskgg2019RealTimeMO} a WaveNet-based model is as an example adapted to predict the current non-linear output sample value, given a certain number of past input samples and the current input. In both works, the number of past input samples, also called the receptive field, is dynamically selected based on the measured impulse response length of the circuits under consideration. In \cite{rnn_vela} a recurrent neural network (RNN), such as the gated recurrent unit (GRU), is proposed for simulating the non-linear behaviour of distortion circuits due to their stateful nature. Contrary to this, the authors of \cite{Bergner} present the state trajectory network (STN), comprised of a standard multilayer perceptron (MLP) with a skip-layer connection surpassing the densely connected layers of the network. The STN differs from related work as the input data is concatenated with measured values from the states of the circuit in order to model its behaviour. Since all aforementioned models are built upon the black-box paradigm, the results are significantly exposed to errors in the data collection process and any flaws in the hardware. Thus both the sonic characteristics and the phase response of the system are learned, introducing arbitrary and non-linear phase shifts to the incoming signal. This becomes a problem where parallel-path processing is desired, for instance when dry-wet mixing with the given simulations. All-pass filters (APF) that have unitary magnitude response and frequency-dependent phase responses would traditionally be the approach to take account of the phase shifts, however, manual coefficient adjustments would be both time-consuming and for specific problems, impossible. An automatic solution to the problem, therefore, is highly desired. With inspiration from the differentiable digital signal processing (DDSP) methodology \cite{DDSP}, we propose a model that tunes the coefficients of a cascaded APF system. The phase response of different black-box effects is thus automatically approximated, and the adjusted APFs are used to align a dry input signal with the processed, phase-shifted output.
\\
\\
\indent The remainder of this paper is structured as follows: the all-pass optimization problem and related work are introduced in section 2. The construction of the differentiable APFs and their formulas are reviewed in section 3. Our approach and different deep optimization architectures are discussed in section 4. Finally, network evaluations, results, allusions and conclusions are presented in sections 5 and 6.

\section{Background}
\label{sec:background}
DDSP stems from the motivation of generating audio by using a deep learning workflow to predict and extract synthesis parameters for vocoders and subtractive synthesis \cite{DDSP}. However, directly integrating classic signal processing elements into deep learning methods has shown promising results for the control and adjustment of other DSP blocks, including convolution, filters and one-period wave-tables \cite{wavetable}. Specifically, the authors of \cite{Kuznetsov2020DIFFERENTIABLEIF} have demonstrated the use of DDSP in the context of IIR filters, training different filter topologies in a recursive manner to match target frequency responses.  Several other projects have investigated deep learning for IIR filter design, but similar to \cite{Kuznetsov2020DIFFERENTIABLEIF}, the work has solely been focused on learning coefficients for magnitude rather than the phase responses. In \cite{Nercessian2020NEURALPE}, a neural network is applied to carry out parametric equalizer matching using differentiable biquads, whereas they in \cite{Bhattacharya} approximate shelving filter coefficients directly in the difference equation. All of these works carry out an optimization problem in the frequency domain by minimizing the mean squared error (MSE) between the ground truth and the derived magnitude responses. \\
\\
\indent The approach to frequency response matching is different in \cite{pepe2021deep}. Here a BiasNet is applied to determine the IIR equalizer parameters. The BiasNet is a simple feedforward neural network that takes advantage of the learnable bias terms, denoted $\textbf{b}_0$, in the input layer. This architecture is called a "deep optimization" algorithm, owing the name to the use of the neural network as a non-convex optimization algorithm used to tune or derive external parameters. An advantage of the BiasNet is its independence of input features, which according to \cite{pepe2021deep} is more likely to provide a solution to many optimization problems. Furthermore, the network does not rely on the input size and content, hence only a target frequency response is required to be given to the loss function. Similar to the IIR system in \cite{pepe2021deep}, adjusting cascaded APFs might be a highly non-linear process. In this paper we, therefore, utilise the over-parameterised nature of the BiasNet to overcome the, potentially, non-convex phase response matching problem and extend the work of \cite{Kuznetsov2020DIFFERENTIABLEIF} to be applicable in the domain of APFs and phase response approximation. 

\subsection{Problem Formulation}
We represent the monophonic signals we want to phase compensate as input vectors $\textbf{x}^T\in \mathbb{R}$, where $T$ is the signal length. The task is to process these signals with an APF function $f$, such that the signal is phase shifted to match a target signal introducing the least amount of destructive interference. The function $f$ takes as arguments the input and the number of filter coefficients $\textbf{c}$ matching the filter order $N$ of the given sub-system. This yields the output $y^T = f(\textbf{c}^n, \textbf{x}^T)$. For a system of cascaded APFs, we define the function composition of size $D$, where each function receives the output of the previous one as:
\begin{equation}
    y^T = f(\textbf{c}^n, \textbf{x}^T)_1 \circ f(\textbf{c}^n)_2 \circ ... \circ f(\textbf{c}^n)_{D-1} \circ f(\textbf{c}^n)_{D},
\end{equation}

\noindent where the order $N = nD$, if each sub-system is a 2nd order filter. The system can be more general than that depending on the value of $n$. Depending on the deep learning techniques used, each function $f$ can be arbitrarily complex and represented either directly as filter coefficients, as done in \cite{Kuznetsov2020DIFFERENTIABLEIF}, or as parameterised sub-networks such as the BiasNet applied for the deep filter optimization procedure in this paper.

\section{Differentiable All-pass Filters}
Before outlining the model architecture of the proposed APF filter tuning process, we present the differentiable APF structures used to adjust the coefficients in the deep learning pipeline. Following the transposed direct form-II (TDF-II) structure, a 2nd order IIR APF is given by the traditional biquad transfer function \cite{dafx}:
\begin{equation}
    A_2(z) = \frac{c + dz^{-1} + z^{-2}}{1 + dz^{-1} + cz^{-2}}
\end{equation} 

In practice, this transfer function can be implemented using the
following recurrent, and stateful, difference equation:

\begin{equation}
\begin{split}
y[n] & = cx[n] + v_1[n] \\
v_1[n] & = dx[n] + v_2[n] - dy[n] \\
v_2[n] & = x[n] - cy[n], 
\end{split}
\end{equation}where coefficients $c$ and $d$ are controlling the steepness and the break frequency of the APF's phase response respectively. The coefficients are products of the pole radius $R$ and the cutoff frequency $f_c$. They have the ranges: $-1 < c < 1$ and $-2 < d < 2$. 
We introduce a stability constraint to the tuning process and estimate the filter parameters rather than the coefficients themselves. The filter coefficients can for each forward call thereafter be calculated by \cite{smithAFX}:
\begin{equation}
    c = R^2 \qquad d = -2R \cos(2\pi f_c/f_s),
\end{equation}with $f_s$ being the sampling rate of the signal. As the filter coefficients, and thus the steepness of the phase response, are dependent on the pole radius, the phase response might have a significantly narrow resolution at low frequencies. When trying to match frequencies below 100 Hz, the parameters for a system with a high sampling rate (192 kHz) exist in very small ranges with $0.9 < R < 0.999$ and the resulting coefficient \textit{d} being between $-1.97 < d < -1.999$, depending on the cutoff frequency. It is hypothesised that the prediction of values in such small ranges might introduce numerical overflow and coefficient quantization errors while being difficult to generalise. We, therefore, propose a differentiable warped all-pass structure to increase the frequency resolution in low-frequency ranges, emphasising the importance of low-frequency content in the learning process. A warped APF is designed and realized on a warped frequency scale. It is achieved by replacing the unit delays of a traditional APF with auxiliary 1st order APFs, whose phase response is used to skew the frequency axis \cite{warp1}. A warped version of the APF in equation (3) is given by the difference equation:
\begin{equation}
    \begin{gathered}
    y[n] = \frac{x[n] (c + a^2 + ad) + v_1[n]}{1 + a^2c + ad}\\
    v_1[n] = x[n](2a + d + ac) + y[n]\big(-2ac - d - a\big) - a^3(x[n] - cy[n]) - v_2[n](a^2 + 1) \\
    v_2[n] = x[n] - cy[n] - \big(a^2(x[n] - cy[n]) + av_2[n]\big),
    \end{gathered}
\end{equation}with $a$ being the warping factor i.e. the coefficient of the inserted auxiliary 1st-order APFs. For stability reasons the warping factor for all inserted APFs is identical and thus gathered into a global, but learnable, variable $a$.

\section{Proposed Method}

\begin{figure*}[ht]
\center
\subfloat[\centering Sequential BiasNet Structure]{{\includegraphics[width=0.95\linewidth]{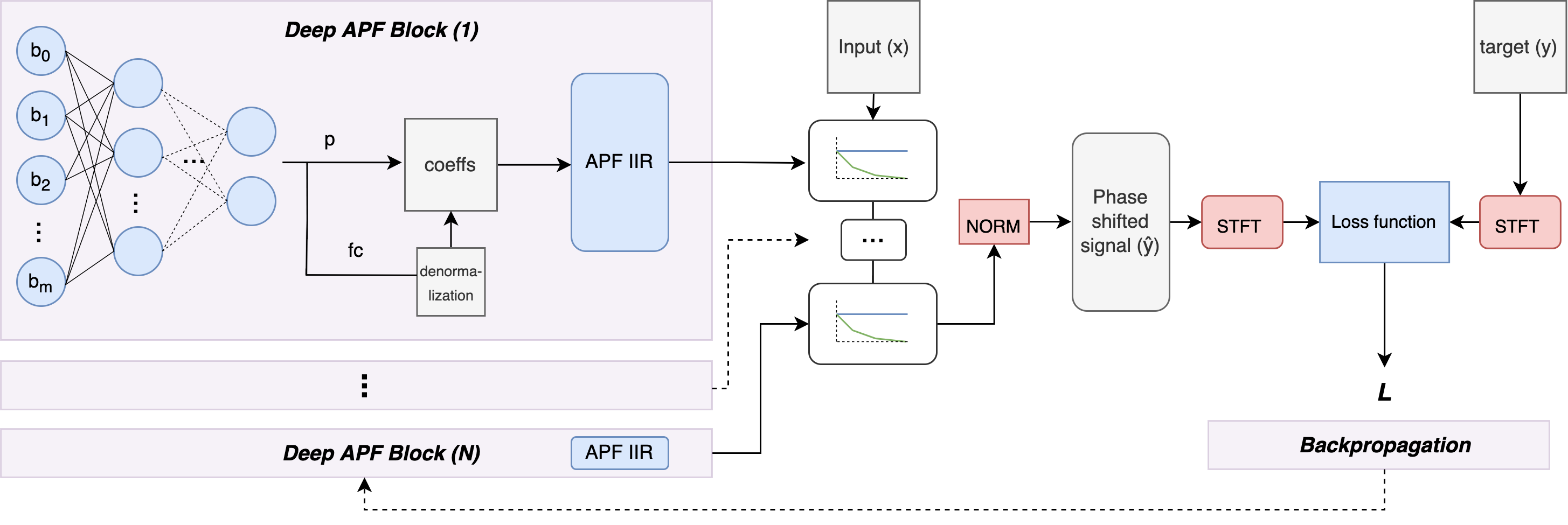} }}%
\qquad
\subfloat[\centering Fully Connected BiasNet Structure]{{\includegraphics[width=0.95\linewidth]{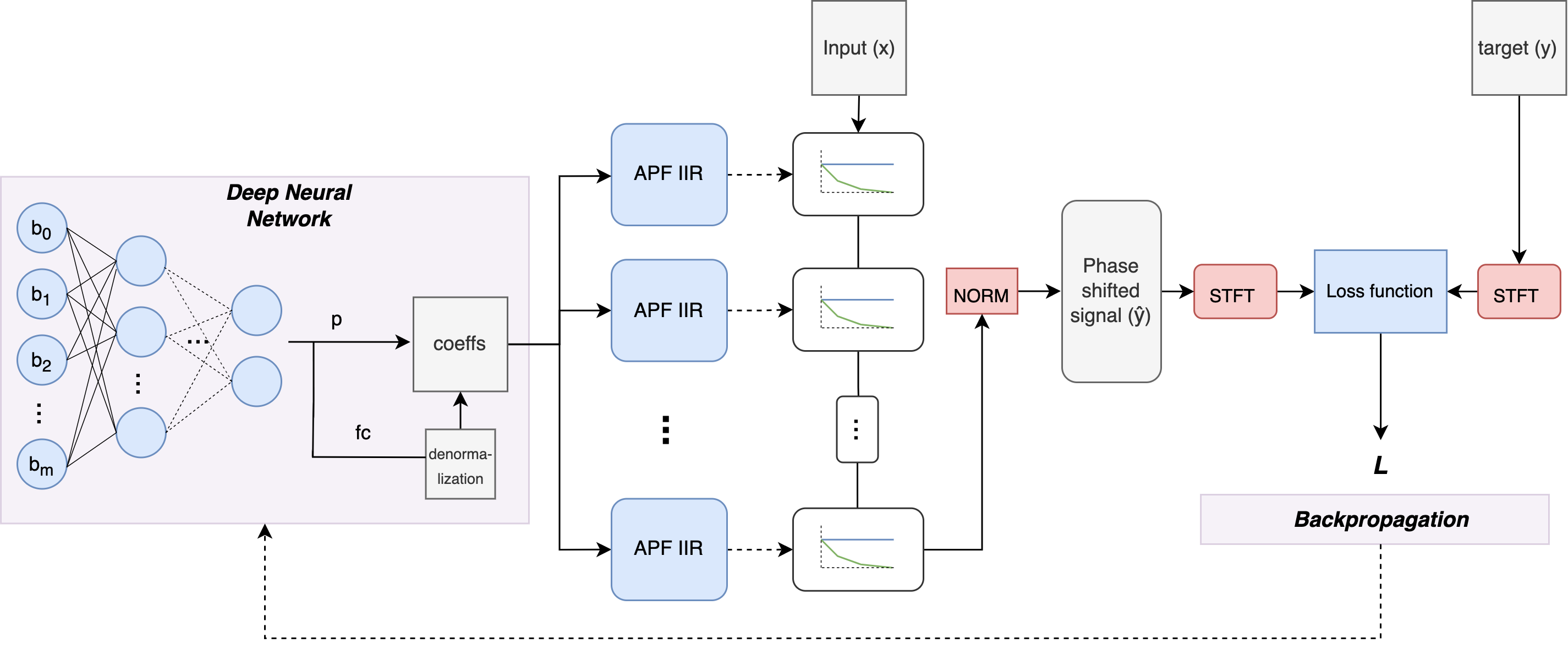} }}%
\caption{\textit{High level overview of the deep-optimization models}}%
\label{fig:models}%
\end{figure*}

By utilizing over-parameterisation we propose two BiasNet-based models and a phase alignment procedure to
extend the differentiable IIR filter design techniques towards the APFs. We call these models \textit{sequential}
and \textit{connected}, as illustrated in figure \ref{fig:models}.
Both models contain cascaded differentiable warped APFs matching a desired filter order $N$. The neural network is excited by a learnable bias input layer, whereas its output corresponds to the filter parameters of the closed-form equations used to calculate the final APF coefficients. Three values, $R$, $fc$ and $a$ are thus fed from the output of the model to every single filter. The primary deep neural network is an MLP with periodic sinusoidal activations for the hidden layers and \textit{tanh} activations for the output layer. The \textit{sine} activation function has been included as it avoids local minima during network optimization, is robust towards vanishing gradients and thus suitable for non-convex problems such as the cascaded APF pipeline \cite{sine}. We additionally de-normalise the network outputs taking account of the range in which $fc$ exist. We use a constrained de-normalization technique similar to the one proposed in \cite{pepe2021deep} to de-normalise the \textit{tanh} output layers scaling it between 20 Hz and 20 kHz:
\begin{equation}
    fc = \frac{fc_{max} - fc_{min}}{2} p + \frac{fc_{max} + fc_{min}}{2},
\end{equation}
where $p$ denotes the value to de-normalise. The DNN is updated such that its output layer produces filter coefficients that create the needed phase alignment. We create two different BiasNet models to investigate the importance of over-parameterisation and its impact on the non-convex problem as well as the general learning process. More specifically, the cascade in the \textit{sequential} structure is achieved by chaining several BiasNets together, each representing a respective filter, to create the desired order. Individual DNNs are thus used to derive the coefficients in parallel for each individual APF. The BiasNet is initialized as a densely connected bottleneck with hidden layers of 1024, 512, 256, 128 units respectively. It contains approximately 692.5k learnable parameters, which accumulate to 2.7 million parameters for a cascaded filter of 7th order. Due to its large number of learnable parameters, a benefit of this architecture is the possibility of a complex and detailed parameter estimation process, however, it suffers from longer calculation times and a lack of interaction between the DNNs of each individual block. For the \textit{connected} structure, all filter parameters are coming directly from one large BiasNet. This introduces only 692k learnable parameters in total, independent of the cascaded filter order. The size of the output layer in the \textit{connected} architecture thus equals 11 for a warped APF of 7th order. The \textit{connected} architecture allows for interaction between the cascaded filters since the same network derives all parameters, however, it might suffer from a smaller and therefore less complex parameter space.

\subsection{Loss Function}
Following the majority of related work, the loss function of the network that is optimized during training happens in the frequency domain.  Since the frequency response of an APF by nature is unity gain, a change in magnitude will not be detected and the direct spectrogram comparison used in \cite{DDSP} and \cite{Nercessian2020NEURALPE} is thus not sufficient for the problem at hand. Rather, we calculate the difference between the sum of the target frequency response $y$ and the predicted signal’s frequency response $\hat{y}$ individually, as well as the frequency response of the summed signals before the transform. The loss function is given by: 
\begin{equation}
    \varepsilon_{STFT}(y_i, \hat{y}_i) = \frac{1}{n} \sum_{i=1}^{n} ( (S(y_i) + S(\hat{y}_i)) - S(y_i + \hat{y}_i))^2,
\end{equation}

\noindent where the function $S$ denotes the spectrogram or the squared magnitude of the STFT, simply given as:
\begin{equation}
    S(y_i) = \lvert STFT(y_i) \rvert ^2
\end{equation}

By doing this, the magnitudes of the target and the prediction are forced to be similar, leaving phase as the only changeable factor. Since the magnitude spectrogram $S$ does not include phase information, it highlights frequency areas where the summation of the input and the target introduce destructive interference. Optimization can thus exclusively be achieved by attaining coefficients whose phase response shifts the input such that the magnitude of the signal summation matches the magnitude summation of each individual STFT. To avoid the frequency-dependent trade-off of the STFT and to improve the robustness of the loss function, we extend equation (7) by the multi-resolution STFT (M-STFT) loss \cite{steinmetz2020auraloss}: 

\begin{equation}
    \varepsilon_{M-STFT}(y_i, \hat{y}_i) = \frac{1}{M} \sum_{m=1}^{M} \varepsilon_{STFT}(y, \hat{y}),
\end{equation}with $M$ being different analysis resolutions. Thus the final loss-function is given by an average over the normal STFT loss in eq (7) at different resolutions. By utilizing multiple FFT-lengths and summing the information across the different resolutions, we capture a more realistic representation of the training signals \cite{steinmetz2020auraloss}. The different resolutions are selected according to the STFT parameters presented in \cite{yamamoto}:

\begin{table}[!htb]
  \centering
  \captionsetup{justification=centering}
    \begin{tabular}{ c c c } 
    \hline
    \textbf{FFT-Size} & \textbf{Hop Length} & \textbf{Window Size} \\
    \hline
    512 & 50 & 240 \\ 
    1024 & 120 & 600\\ 
    2048 & 240 & 1200\\ 
    \hline
    \end{tabular}
    \caption{\label{tab:fft_specs}\textit{Details of the parameters for the different STFT resolutions}}
\end{table}

\begin{figure*}[tbph]
    \centering
    \captionsetup{justification=centering}
    \includegraphics[width=0.95\linewidth]{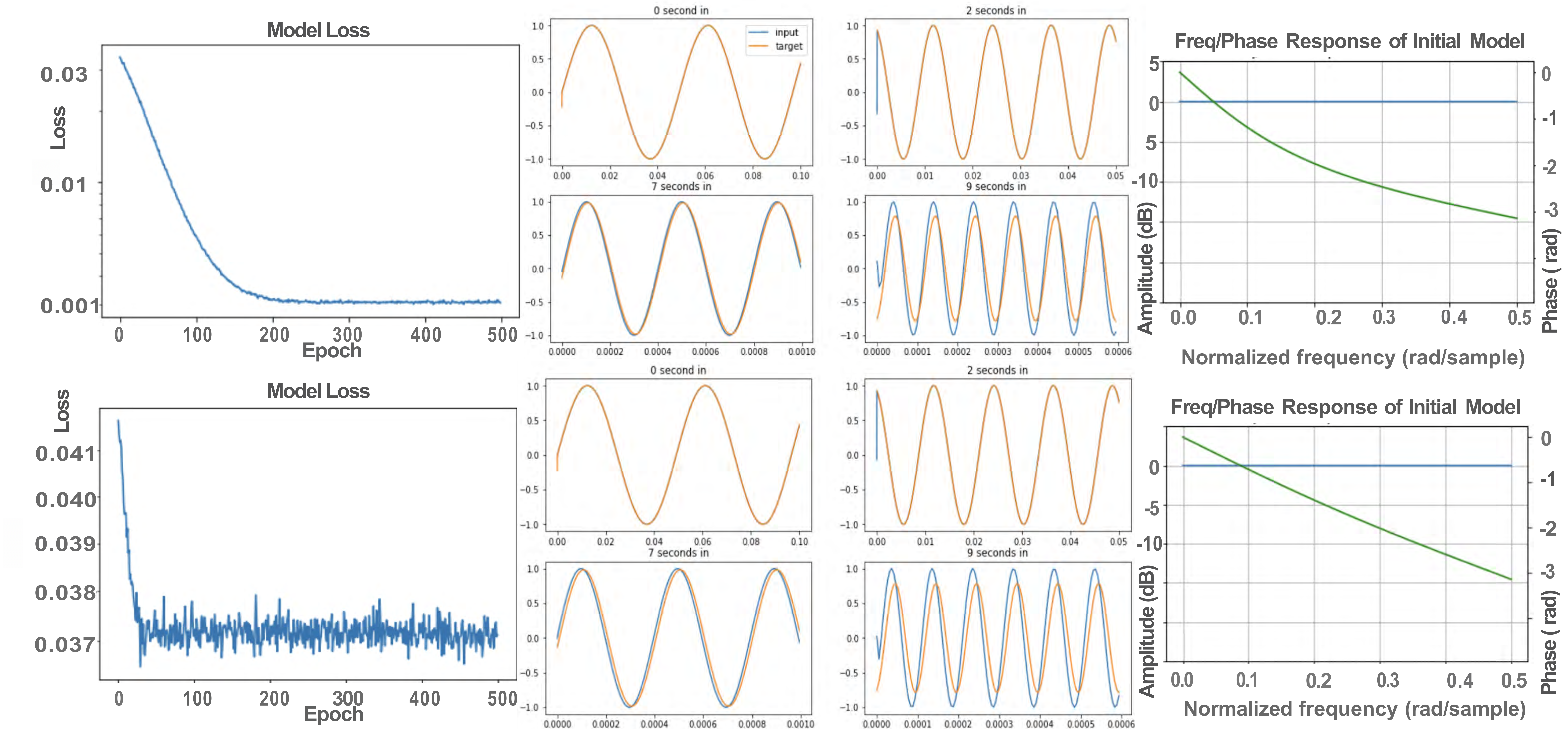}
    \caption{\textit{Loss curve, phase compensation and phase response of RC Filter trained with M-STFT Loss (upper row) and the MSE Loss (lower row). The middle plot shows the alignment at 0, 2, 7 and 9 seconds into the training signal.}}%
    \label{fig:rc_sim}
\end{figure*}

\subsection{Proof of concept}
\sloppy
By a simple proof of concept we show that the over-parameterisation proposed above is crucial for the deep APF optimization problems at hand. To inspect the possibilities of differentiable APFs we first create an example following the work in \cite{Kuznetsov2020DIFFERENTIABLEIF}. We thus start with a naive DDSP approach and derive the coefficient values for the filters directly from the difference equation in order to provide a baseline. To do this we attempt to align the input and output of a simulated 1st-order RC filter, which due to its natural low-pass behaviour creates a phase shift in the higher frequency register. A 1st order RC filter is given by the difference equation \cite{FabianWhiteBox}:
\begin{equation}
    V_{out}^n = \frac{\rho(V_{in}^n + V_{in}^{n-1}) + (1 - \rho) V_{out}^{n-1}}{1 + \rho}, 
\end{equation}where $V_{in}^{n}$ and $V_{in}^{n-1}$ are the current and past input samples, $V_{out}^{n}$ and $V_{out}^{n-1}$ are the current and past output samples and $\rho$ is given by $fs/(2RC)$, with $R$ and $C$ being the resistance and capacitance of the circuit components. In our case $R = 120 \Omega$ and $C = 68 nF$ respectively. As the RC filter at maximum will shift incoming frequencies $90^{\circ}$, we train a 1st-order APF the naive way, using the same hyperparameters and loss functions presented in section \ref{evaluation}. The results of the trainings are depicted in figure \ref{fig:rc_sim}.

As seen above, both the M-STFT and the MSE loss converge, with the latter being faster but more noisy. Both cases additionally manage to compensate the phase shifts introduced by the RC filter, with the M-STFT training being more precise. However, when applying the above naive approach to more complex problems such as the VA black box effects presented in section \ref{evaluation}, it was quickly realized that the training loss for a system of cascaded APFs diverged and in many cases exploded. When tuning cascaded APFs we are simply handling a highly non-linear problem where the individual minimum of each APF affects the remaining cascade, while the minimum of the loss function most often is based on the frequency areas where alignment gives less destructive interference. The function that can estimate the full phase response thus might be non-convex as it has multiple local minima, which was found to be too complex for the naive and traditional DDSP approach. We argue that over-parameterisation networks and deep optimization frameworks solve this problem.
\section{Evaluation}
\label{evaluation}
We examine the proposed models through objective metrics and use the proposals with the best results for final listening tests. The training data consists of a logarithmic sine sweep from 20 Hz to 20 kHz over 10 seconds at a sample rate of 192 kHz. The sweep is fed through three different VA black box simulations shown to introduce significant and complex deviations from linear phase behaviour: 1) Electronic Audio Experiments Surveyor Pre-amp, 2) 15IPS Tape Saturation, and 3) LEM 808R DLX Mixer. We train and evaluate all models in an \textit{in-to-out} fashion, meaning that our models learn the coefficients needed to shift the non-affected input in order to match the VA processed and phase-shifted output. Once training is done, the coefficient values can be exported and inserted into a traditional APF pipeline for the desired real-time adjustments. All models are initialized as a 7th-order APF structure with a cascade of three 2nd-order filters and one 1st-order APF. The output signals of the three systems are sampled and divided into sequences of 2048 samples, which for 20 Hz approximates to a 1/4 of a sinusoidal period at a sample rate of 192 kHz. We heuristically found this sequence length to be a good compromise between phase information and training time. The sub-sequences are additionally organized into batches. We train the models using the earlier mentioned M-STFT loss as well as the traditional MSE loss function, which is used to further validate the phase-compensated simulations/reconstructions. All training sessions are carried out using 1 NVIDIA Tesla T4 GPU for 400 epochs or until training loss plateaus (approx. 5 hours). We train the models with a learning rate of 1e-5, a batch size of 512 and the ADAM optimizer. The final M-STFT and MSE values for the trained models are seen in table \ref{tab:training_details} below:
\begin{table}[!htb]
  \centering
  \captionsetup{justification=centering}
    \begin{tabular}{l c c l|c} 
    \hline
    \textbf{Model} & \textbf{Loss Type} & \textbf{Params} & \textbf{Effect} & \textbf{Final Loss} \\
    \hline
    \multirow{3}{*}{Sequential} & \multirow{3}{*}{MSE} & \multirow{3}{*}{2.7M} & Surveyor & 1.375e-1 \\
    & & & 15 IPS & 4.235e-3 \\
    & & & LEM & 4.708e-2 \\
    \hline
    \multirow{3}{*}{Sequential} & \multirow{3}{*}{M-STFT} & \multirow{3}{*}{2.7M} & Surveyor & 1.857e-2 \\ 
    & & & 15 IPS & 8.078e-3 \\
    & & & LEM & 9.998e-3 \\
    \hline
    \multirow{3}{*}{Connected} & \multirow{3}{*}{MSE} & \multirow{3}{*}{692.5K} & Surveyor & 1.437e-1 \\ 
    & & & 15 IPS & 4.271e-3 \\
    & & & LEM & 4.707e-2 \\
    \hline
    \multirow{3}{*}{Connected} & \multirow{3}{*}{M-STFT} & \multirow{3}{*}{692.5K} & Surveyor & 2.991e-1 \\
    & & & 15 IPS & 4.871e-1 \\
    & & & LEM & 9.887e-3 \\
    \hline
    \end{tabular}
    \caption{\label{tab:training_details}\textit{Model and training specifications}}
\end{table}

\begin{figure*}[ht]
    \centering
    \captionsetup{justification=centering}
    \includegraphics[width=1.0\linewidth]{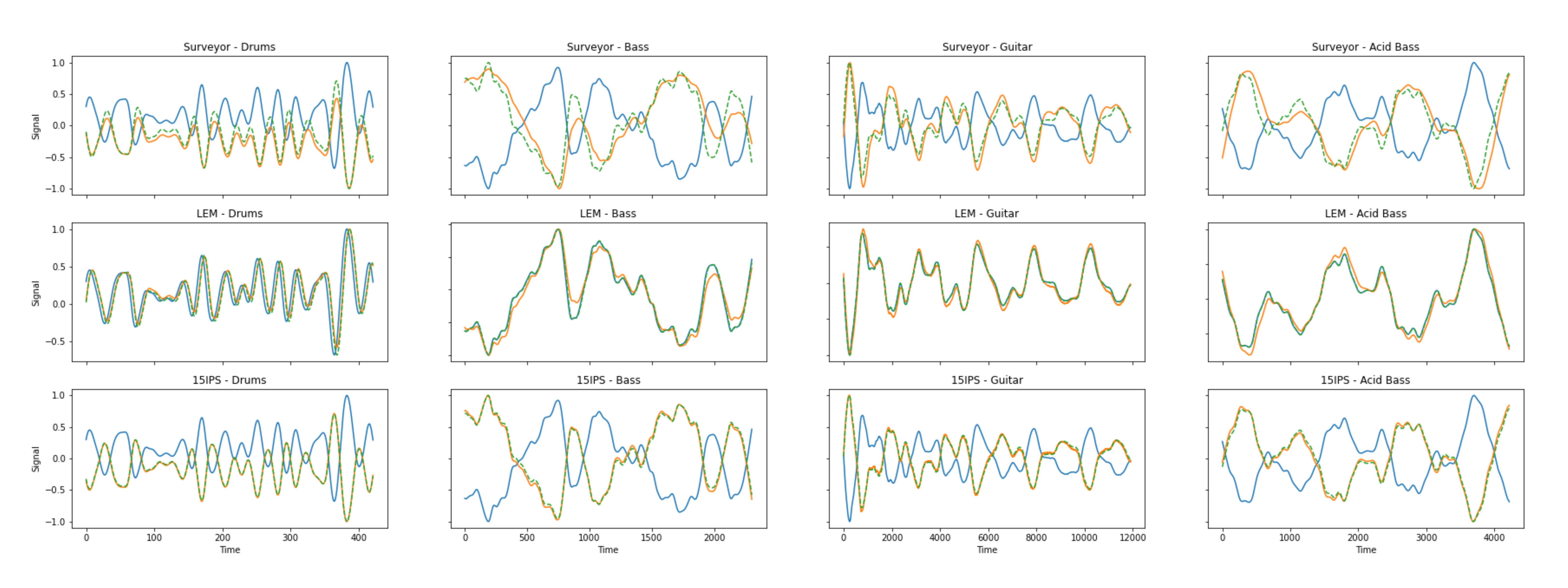}
    \caption{\textit{Examples of the normalized phase alignment results for each individual black-box effect training (blue = input, orange = output, green = prediction}}%
    \label{fig:phase}
\end{figure*}
\subsection{Performance Assessment}
The performance of the trained models is quantitatively evaluated on unseen test audio. The test audio is chosen such that it exposes the model to signals of various frequencies and timbres. It consists of a concatenation of different loops counting: an acoustic
breakbeat drum loop, an electric bass-guitar loop, a guitar loop and a synthesized acid bassline (duration of approx 2 minutes). As signal displacement is given in the time domain, the objective metrics chosen for this study compare the similarity between the automatically shifted input (prediction) and the VA processed output (ground truth) in the sample/phase space. We evaluate the performance by measuring the similarity using the traditional MSE as well as the mean absolute error (MAE) defined as: 
\begin{equation}
    \epsilon_{MAE}(\hat{y}, y) = \frac{1}{n-1} \sum_{i=0}^{n-1} \lvert \hat{y_i} - y_i\rvert
\end{equation}

Additionally, we include the error-to-signal ratio (ESR), which can be regarded as an extension of the MSE with the inclusion of target energy normalization to penalise the errors more equally when the input signal is lower in absolute amplitude. The ESR is given by:

\begin{equation}
    \epsilon_{ESR}(\hat{y}, y) = \frac{\sum_{i=0}^{N-1} \lvert \hat{y_i} - y_i\rvert^2}{\sum_{i=0}^{N-1} \lvert y_i \rvert^2}
\end{equation}
\\
The final values for each objective metric are summarized in table \ref{tab:dl_results}, with the values for the non-shifted signals included as a static reference. We see that the \textit{sequential} architecture performs best for all objective metrics on both the \textit{surveyor} and the \textit{15IPS} effects (given in bold). The \textit{sequential} model is slightly surpassed by the \textit{connected} architecture in the case of the \textit{LEM} effect, however, only with a combined distance of 0.003 for the MSE trained version and 0.001 for the M-STFT trained version. It can thus be concluded that the \textit{sequential} and over-parameterised BiasNet approach quantitatively provides a closer match to the ground truth phase response of the trained systems. Figure \ref{fig:phase} presents a few phase-matching results on different frequency content of the test audio. Examples of all trained VA simulations are shown. It is here clearly seen that the phase-response estimation done by the models compensates for the input to match the saturated and phase-shifted output.

\begin{table}[!htb]
  \centering
  \captionsetup{justification=centering}
    \begin{tabular}{ l c l|c|c|c } 
    \hline
    \textbf{Model} & \textbf{Loss} & \textbf{Effect} & \textbf{MAE} & \textbf{MSE} & \textbf{ESR} \\
    \hline
    \multirow{3}{*}{Reference} & \multirow{3}{*}{None} & Surveyor & 0.161 & 0.066 & 3.942 \\
    & & 15 IPS & 0.164 & 0.069 & 4.105 \\
    & & LEM & 0.020 & 0.001 & 0.068 \\
    \hline
    \multirow{3}{*}{Sequential} & \multirow{3}{*}{MSE} & Surveyor & \textbf{\textit{0.033}} & \textbf{\textit{0.003}} & \textbf{\textit{0.177}} \\ 
    & & 15 IPS & \textbf{\textit{0.007}} & \textbf{\textit{0.0005}} & \textbf{\textit{0.007}} \\
    & & LEM & 0.017 & 0.0007 & 0.045 \\
    \hline
    \multirow{3}{*}{Sequential} & \multirow{3}{*}{MSTFT} & Surveyor & 0.037 & 0.004 & 0.235 \\ 
    & & 15 IPS & 0.015 & 0.001 & 0.058 \\
    & & LEM & 0.017 & 0.0007 & 0.045 \\
    \hline
    \multirow{3}{*}{Connected} & \multirow{3}{*}{MSE} & Surveyor & 0.079 & 0.017 & 1.022 \\ 
    & & 15 IPS & 0.010 &  0.0002 & 0.017 \\
    & & LEM & \textbf{\textit{0.016}} & \textbf{\textit{0.0007}} & \textbf{\textit{0.042}} \\
    \hline
    \multirow{3}{*}{Connected} & \multirow{3}{*}{MSTFT} & Surveyor & 0.074 & 0.015 & 0.888 \\
    & & 15 IPS & 0.089 & 0.021 & 1.236 \\
    & & LEM & 0.017 & 0.0007 & 0.044 \\
    \hline
    \end{tabular}
\caption{\label{tab:dl_results}\textit{Overview of the performance results for the individual models across effects and loss functions}}
\end{table}

\subsection{Listening Test}
Due to the inadequacy of the objective metrics in evaluating the perceived quality of the phase alignment in real life use-cases, such as parallel path processing scenarios, a subjective listening
test is carried out. By the use of an 'audio perceptual evaluation' (APE) listening test, we examine the difference between the clean summation and the compensated dry-wet mixing of musical content, using the proposed \textit{sequential} architecture. The APE style test extends the well-known MUSHRA test by rating different versions of the same reference on a single scale using sliders \cite{APE}. Compared to the MUSHRA test, the APE is useful for evaluating the perceived quality of dry-wet mixing as there exists no known reference. Since the audibility of dry-wet mixing highly differs relative to the use case, the participants are presented with three different musical scenarios for each compensated audio effect: a low relative mix with 75\% dry signal and 25\% wet signal, a middle relative mix with 50\% dry signal and 50\% wet signal, and a high relative mix
with 25\% dry signal and 75\% wet signal. Each relative mix is normalized, however, no loudness compensation has been applied as volume differences in different frequency areas are natural artifacts of phase misalignment and thus represents the baseline of the listening test. We present the participants with two different audio mixes matching a real-world music mixing and mastering scenario, where the black-box effect would be applied to give the final mix a saturating "warmth". The participants are informed that they are listening to different versions of effect models and thereafter instructed to ’blindly’ compare the clean and compensated versions based on their perceived level of audio quality. Sound examples can be heard on the accompanying webpage \footnote{https://abargum.github.io/}.

\begin{figure*}[!pbt]
    \centering
    \captionsetup{justification=centering}
    \includegraphics[width=0.9\linewidth, height=0.32\linewidth]{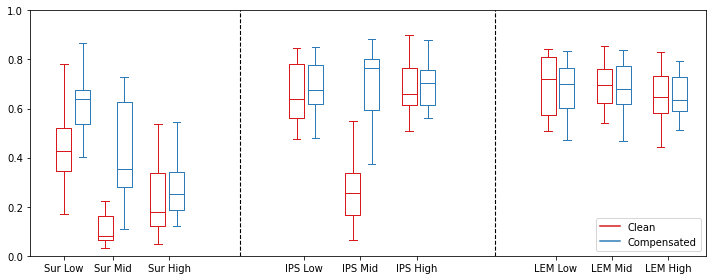}
    \caption{\textit{Ratings across each individual mix-case}}%
    \label{fig:boxplot_2}
\end{figure*}

\indent 15 convenience sampled participants without any reported hearing impairments and 3 or more years of musical experience took part in an online listening test. Individual boxplots for the evaluation of the clean and compensated audio mixes are shown in figure \ref{fig:boxplot_2}. The answers for each audio mix are summed and averaged for each participant, giving a final comparable score for the individual black-box effects across the different mix configurations.

\indent As seen in \ref{fig:boxplot_2}, the difference between the clean and compensated versions for the ’middle’ scenario with 50\% dry-wet mixing is highly audible. This is evident both for the \textit{Surveyor} and the \textit{15IPS} effects. The ’low’ mix scenario additionally performs better for the compensated version for both the surveyor and the 15IPS. In the case of the LEM effect, all dry/wet mix cases were rated to sound equally good. As seen in figure \ref{fig:phase} this is most likely caused by the lack of phase shifts happening in the audible frequency ranges. Lastly, the scores for all the ’high’ cases barely differed, which possibly is due to the fact that the dynamics of the saturated output masks the actual interference.

It is thus clear that the trained models manage to align the input to its respective target signal in the presented examples. This is quantitatively evident in the objective metrics in table 3 where the ’sequential’ MSE model perform better on all metrics, compared to its static counterpart. The time-domain representation in figure 3, furthermore, supports the alignment of the musical signals where it clearly can be seen that the temporal envelope of the prediction matches the target. Lastly, a perceptual listening test show that especially the audio quality of the surveyor and the 15IPS models are improved in the dry-wet mixes provided.

\section{Conclusions}
\sloppy
To address the challenges of the learned phase responses in VA black box effects, this paper has presented, discussed and evaluated deep-learning techniques for automatic signal alignment. By utilizing the ’deep optimization’ methodology, we propose a BiasNet-inspired architecture that approximates filter parameters used for coefficient calculations in a system of cascaded differentiable warped APFs. We thus extend the naive approach to approximating DDSP IIR filters with over-parameterized neural networks and use them to exhibit successful models for aligning the dry and wet paths of virtual analog effects. Ultimately, three black-box effects are chosen for the final training procedure. By evaluating the models on different objective metrics, we demonstrate that what we call a 'sequential' architecture efficiently tunes all-pass filter coefficients for approximating a system's phase response. It is thus demonstrated that over-parameterisation is suitable when estimating filter coefficients in more complex and non-convex scenarios. The results are supported by subjective listening tests, where 15 expert listeners rated the dry-wet mixing of VA effects to be significantly improved by the deep all-pass models, proving that the approach additionally is useful in real life use-cases.

\section{Acknowledgments}
Many thanks to the great number of anonymous reviewers and the whole ML/DSP research team at Native Instruments Berlin in 2022 for their help, guidance and support.

\nocite{*}
\bibliographystyle{IEEEbib}
\bibliography{DAFx23_tmpl} 
\end{document}